
\input tables.tex
\input phyzzx.tex

\def\tauptaum{\tau^+\tau^-}

\def\chitil{\widetilde\chi}
\def\cnone{\chitil^0_1}
\def\hsm{H}
\def\mhsm{m_{\hsm}}

\def\chitil{\widetilde\chi}

\def\cnone{\chitil^0_1}

\def\wpm{W^{\pm}}
\def\wp{W^+}
\def\wm{W^-}
\def\hl{h^0}
\def\mhl{m_{\hl}}

\def\mt{m_t}

\def\hl{h^0}

\def\mhl{m_{\hl}}

\def\gam{\gamma}
\def\gam{\gamma}

\def\gam{\gamma}

\def\ptmiss{p_T^{\,\,miss}}

\def\gev{~{\rm GeV}}
\def\tev{~{\rm TeV}}

\def\fbi{~{\rm fb}^{-1}}


\def\prdj#1{{\it Phys. Rev.} {\bf D{#1}}}
\def\npbj#1{{\it Nucl. Phys.} {\bf B{#1}}}
\def\prlj#1{{\it Phys. Rev. Lett.} {\bf {#1}}}
\def\plbj#1{{\it Phys. Lett.} {\bf B{#1}}}

\def\mt{m_t}
\def\wp{W^+}
\def\wm{W^-}
\def\rta{\rightarrow}

\def\nsd{N_{SD}}

\def\mw{m_W}
\def\mz{m_Z}
\def\anti{\overline}

\def\ifmath#1{\relax\ifmmode #1\else $#1$\fi}

\def\3quarter{{\textstyle{3 \over 4}}}

\input phyzzx
\Pubnum={$\caps UCD-94-10$\cr}
\date{April, 1994}

\titlepage
\vskip 0.75in
\baselineskip 0pt
\hsize=6.5in
\vsize=8.5in
\centerline{{\bf Standard Model Higgs Physics at a 4 TeV
Upgraded Tevatron}}
\vskip .075in
\centerline{ J.F. Gunion and T. Han}
\vskip .075in
\centerline{\it Davis Institute for High Energy Physics, Dept. of Physics}
\centerline{\it University of California--Davis, Davis, CA 95616}

\vskip .075in
\centerline{\bf Abstract}
\vskip .075in
\centerline{\Tenpoint\baselineskip=12pt
\vbox{\hsize=12.4cm
\noindent We compute an array of Standard Model Higgs boson ($\hsm$)
signals and backgrounds
for a possible upgrade of the Tevatron to $\sqrt s=4\tev$.
Taking $\mt\geq 140\gev$, and assuming a total accumulated luminosity of
$L=30\fbi$, we find that a Standard Model Higgs boson with $\mhsm\lsim
110\gev$ could almost certainly be detected using the
$\wpm\hsm\rta l\nu b\anti b$ mode.
A Higgs boson with mass between $\sim 120\gev$ and $\sim 140\gev$
or above $\sim 230-250\gev$ almost certainly would not be seen.
A Higgs boson with $\mhsm\sim 150\gev$ or $200\lsim\mhsm\lsim 230-250\gev$
has a decent chance of being detected in the $ZZ\rta 4l$ mode.
There would also be some possibility of discovering the $\hsm$
in the $WW\rta l\nu jj$ mode for $150\lsim\mhsm\lsim 200\gev$.
Finally, hints of an event excess in the $WW\rta ll \nu\nu$ mode due
to the $\hsm$ might emerge for $140\lsim\mhsm\lsim 180\gev$.
Given the difficult nature of the Higgs boson signals for $\mhsm$
values beyond the reach of LEP-200, and the discontinuous $\mhsm$
range that could potentially be probed, justification of
an upgrade of the Tevatron to $4\tev$ on the basis of its potential
for Standard Model Higgs boson discovery would seem inappropriate.
}}

\vskip .15in
\noindent{\bf 1. Introduction}
\vskip .075in

Given the cancellation of the SSC, it is important to reassess the
possibilities for exploring the Electroweak Symmetry Breaking (EWSB)
sector at possibly available machines.  Here we focus on the
Standard Model (SM) Higgs boson ($\hsm$). It is well-known that
the LHC can find the $\hsm$ for all masses between about $80\gev$
and $\sim 800\gev$,\Ref\arubbia{For a recent review, see {\it e. g.},
A. Rubbia,  p. 601, in {\it Physics and Experiments with
Linear $e^+e^-$ Colliders},
Waikoloa, Hawaii, eds. F. A. Harris {\it et al.}, (1993).}\
and possibly explore a strongly interacting EWSB
sector that would arise for very large effective $\mhsm$ values.
\Ref\baggeretal{J. Bagger, V. Barger, K. Cheung, J.F. Gunion, T. Han,
G. Ladinsky, R. Rosenfeld, and C.P. Yuan, \prdj{49} (1994) 1246;
M. Chanowitz and W. Kilgore, LBL-34849 (1994).}\
However, the time scale for construction of the LHC may be quite
long without significant U.S. participation.  The question arises as
to whether the U.S. should consider an alternative investment of the
same money in existing U.S. laboratories. A Tevatron upgrade to the
$p \anti p$ center
of mass energy $\sqrt s=4\tev$ with yearly luminosity of $L=10\fbi$ has
emerged as a subject of discussion in this context.
\Ref\chill{C. Hill, private discussions; G. Jackson, talk presented at
the Workshop on Electroweak Symmetry Breaking and TeV-Scale Physics,
UC--Santa Barbara, Feb., 1994.}
The possibility of a $pp$ collider with $L=100\fbi$ is even being
considered. Here, we present a first-level examination of
the ability of such upgraded Tevatrons (UT's) to probe
the SM Higgs sector. In particular, we wish to
establish the extent to which an upgraded Tevatron
can search for Higgs bosons with
mass beyond the reach of LEP-200 (\ie\ LEP-II
operated at $\sqrt s=200\gev$).
We compute signals and those backgrounds
that are not highly detector-dependent for all
conceivably useful channels. In some cases, we compare results for a detector
similar to the current CDF and D0 detectors to those for
a more optimized detector.

Current LEP and Tevatron data are on the verge of playing
a significant role in placing limits on the allowed range of $\mhsm$
in the SM.  For instance, should $\mt$ be of order $170\gev$,
fits to the precision electroweak data from LEP and elsewhere
imply that $\mhsm\gsim 150\gev$ in order to be less than
two standard deviations away from the best fit value of $\mhsm\sim 800\gev$.
\Ref\lepcontours{V. A. Novikov {\it et al.}, CERN-TH.7137/94, and references
therein.}\ While this indication of a heavy Higgs boson
in the SM is clearly quite preliminary
at this point in time, a determination at the Tevatron of $\mt$,
coupled with still more precise LEP data, may well pinpoint
a favored region for $\mhsm$.

We now list the discovery channels for the $\hsm$ that we have considered.
For those modes that contain charged leptons ($l$)
we retain only events with $l=e,\mu$.
The production/decay  modes of interest
in rough order of utility are the following:
\pointbegin
Associated $W^\pm\hsm$ production followed by $W\rta l\nu$ and
$\hsm\rta b\anti b$, leading to an $lb\anti b$ final state.
This final state has been considered in the context of the SSC/LHC
(where $t\anti t\hsm$ production with $t\rta W b$ is the dominant
source) in
\REF\daietal{J. Dai, J.F. Gunion and R. Vega, \prlj{71} (1993) 2699.}
\REF\willenetal{A. Stange, W. Marciano, and S. Willenbrock,
\prdj{49} (1994) 1354.}
Ref.~[\daietal] and for the Tevatron (where $W^*\rta W\hsm$
dominates) in Ref.~[\willenetal]. For $W^*\rta W\hsm$,
single or double $b$-tagging is employed to isolate
the final state of interest. However,
even with $b$-tagging, the $W^\pm j j$ background is significant,
and the $W^\pm b\anti b$ and $W^\pm Z\rta W^\pm b\anti b$ processes
are irreducibly present.
\point
Associated $Z\hsm$ production followed by $Z\rta 2l$ and $\hsm\rta b\anti b$.
Backgrounds are the $Z$ versions of the ones noted above.\refmark{\willenetal}
\point
Inclusive $\hsm$ production followed by $\hsm\rta ZZ^*\rta 4l$ decay.
It is well established
\REF\gkw{J.F. Gunion, G. Kane, and J. Wudka, \npbj{299} (1988) 231.}
\REF\sdctdr{SDC Technical Design Report, SDC-92-201.}
\REF\cmsatlas{CMS Letter of Intent, CERN/LHCC 92-3;  ATLAS Letter of
Intent, CERN/LHCC 92-4.}
\refmark{\gkw,\sdctdr,\cmsatlas}\
that there is no sizeable background to this channel for $\mhsm<2\mz$. The
only significant background for $\mhsm>2\mz$ is from the $ZZ\rta 4l$ continuum.
However, the signal rates are low.\Ref\zzllll{This process was considered
for an upgraded Tevatron of 3.6 TeV by V. Barger and T. Han, p. 116 in
{\it Physics at Fermilab
in the 1990's}, Breckenridge, CO, eds. D. Green and H. Lubatti (1990).}\
\point
Inclusive $\hsm$ production followed by $\hsm\rta \wp\wm \rta l\nu jj$.
Backgrounds include the mixed QCD/EW $\wpm jj$ processes, the $\wp\wm$
continuum, and $t\anti t\rta \wp\wm X$.  However, this latter
background is easily eliminated by vetoing against extra jet activity.
\point
Inclusive $\hsm$ production followed by $\hsm\rta \wp\wm\rta 2l 2\nu$.
Assuming we stay away from $2l$ masses in the vicinity of $\mz$,
and veto events with significant jet activity in the central region
(such as those that might come from $t\anti t$
production\Ref\ttveto{V. Barger, G. Bhattacharya,
T. Han, and B. Kniehl, \prdj{43} (1991) 779.}),
the only large backgrounds are from continuum $\wp\wm$
and $\tauptaum$ production. The latter background can be reduced
to a negligible level by an appropriate cut on the transverse-plane angle
between the two final leptons.
\point
Inclusive $\hsm$ production followed by $\hsm\rta ZZ\rta ll \nu\nu$.
The $ZZ\rta ll\nu\nu$ continuum background is certainly present
(and will be the only one considered here), but there are
other detector-dependent backgrounds that could be large --- \eg\
$Zg$ production with $g$ yielding little visible energy.  This
latter background can be eliminated at high $ZZ$ invariant mass
by requiring very small transverse hadronic energy in association
with the two leptons,
\REF\zgelimination{R. N. Cahn {\it at al.}, p. 20 in {\it Proceedings of
Experiments, Detectors, And Experimental Areas for the Supercollider},
eds. R. Donaldson and M. G. D. Gilchriese, Berkeley, (1987).}
\refmark{\zgelimination,\sdctdr}
but no studies have been performed at the low $ZZ$ masses
of relevance here.
\point
Inclusive $\hsm$ production followed by $\hsm\rta ZZ\rta 2l jj$.
Backgrounds include the mixed QCD/EW $Z jj$ processes and the
$ZZ$ continuum.
\point
Inclusive $\hsm$ production followed by $\hsm\rta \gam\gam$ decay.
The primary background for a detector with excellent jet/photon
discrimination power is the irreducible $q\anti q\rta \gam\gam$
continuum.
\point
Inclusive $\hsm$ production followed by
$\hsm\rta \tau^+\tau^- \rta 2l 4\nu$ decay.
The primary backgrounds are the Drell-Yan processes
$\gam^*, Z \rta \tau^+\tau^-$ and $l \anti l$, and
$\wp\wm\rta 2l 2\nu$.

Of these processes, we find that only four have any real chance
of being useful. Assuming a total accumulated luminosity of
$L=30\fbi$, we will see that the $\wpm\hsm\rta l\nu b\anti b$ channel,
with single or double $b$-tagging,
can be used to detect a Standard Model Higgs boson for $\mhsm\lsim
110\gev-120\gev$. However, other modes must be considered at higher mass.
A Higgs boson with $\mhsm\sim 150\gev$ or $200\lsim\mhsm\lsim 230\gev$
has a decent chance of being detected in the rather clean,
but event-rate-limited, $ZZ\rta 4l$ mode.
There would also be some possibility of discovering the $\hsm$
in the $WW\rta l\nu jj$ mode for $150\lsim\mhsm\lsim 200\gev$.
However, to do so requires detection of a $\sim 30\gev$ wide mass peak
over a broadly-peaked background that is 50 to 100 times larger.
In the $WW\rta 2l 2\nu$ mode, the $\hsm$ leads to a broad 10\% to 20\% event
excess in the di-lepton mass distribution for $140\lsim\mhsm\lsim 180\gev$.
But, the signal and $WW$ continuum background have very similar shapes.
Since systematic uncertainties in the background normalization are
unlikely to be brought much below the 10\% level, this channel
will probably at best provide only a hint of the presence of an
$\hsm$ in this mass range.
Regarding the $ZZ\rta ll \nu\nu$ mode, nominal $S/\sqrt B$
values including only the $ZZ$ continuum background are encouraging
in the mass range $200\lsim\mhsm\lsim 230-250\gev$,
but we are unable to draw any final conclusions without further
study of the very severe $gZ$ and related backgrounds,
that could easily overwhelm the signal, depending upon the
precise machine and detector design.
A Higgs boson with mass between $\sim 120\gev$ and $\sim 140\gev$
or above $\sim 230-250\gev$ almost certainly would not be seen.
Our results are insensitive to $\mt$ for values of $\mt\gsim140\gev$.

At times we will quote $S/\sqrt B$ values for a given channel
as an indication of the absolute best that one can achieve in the
absence of systematic effects and/or additional backgrounds.
The reader is warned to pay close attention to the comments
associated with each channel, as for some channels these nominal
$S/\sqrt B$ values are far higher than will be achieved in reality,
serving only to indicate an initial `starting point' before
including all additional effects. In the absence of systematic
effects, we regard $S/\sqrt B\sim 5$ as an appropriate criterion
for discovery.

\FIG\xsec{
Total cross sections for SM Higgs boson production  (without cuts)
are presented versus $\mhsm$ at the upgraded Tevatron
for the major reactions of interest.}
\midinsert
\vbox{\phantom{0}\vskip 5.0in
\phantom{0}
\vskip .5in
\hskip -0pt
\special{ insert scr:hxsec_4tev.ps}
\vskip -1.45in }
\centerline{\vbox{\hsize=12.4cm
\Tenpoint
\baselineskip=12pt
\noindent
Figure~\xsec:
Total cross sections for SM Higgs boson production  (without cuts)
are presented versus $\mhsm$ at the upgraded Tevatron
for the major reactions of interest.
}}
\endinsert

Before proceeding to our detailed results, it is useful to first
present the total cross sections for the various production reactions of
interest.  These appear in Fig.~\xsec.
In this figure, as well as later graphs, we have
included  $K$ factors of 1.5 and 1.2 for the $gg\rta \hsm$
\Ref\djouadi{A. Djouadi, M. Spira, and P. M. Zerwas,
\plbj{264} (1991) 441; S. Dawson, \npbj{368} (1991) 283;
D. Graudenz, M. Spira, and P. M. Zerwas, \prlj{70} (1993) 1372.}\
and $\wpm\hsm$
(and $Z\hsm$)\Ref\hanohn{T. Han and S. Willenbrock,
\plbj{273} (1990) 167; J. Ohnemus and W. J. Stirling, \prdj{47}
(1993) 2722; H. Baer, B. Bailey, and J. Owns, \prdj{47} (1993) 2730.}\
associated production processes, respectively. We have chosen to use the
D0' set of parton distribution functions by MRS.\Ref\mrspdf{A.D. Martin,
R.G. Roberts, and W.J. Stirling, \prdj{47}, (1993) 867.}
The most important point to note from Fig.~\xsec\ is the fact
that in the low $\mhsm$ region of interest the $gg$ fusion
cross section is relatively independent of $\mt$ once $\mt\gsim140\gev$,
\ie\ once $\mhsm$ is substantially below $2\mt$. (A curve for $\mt=80\gev$
is shown to illustrate the much larger cross sections that were
anticipated for $\hsm$ production in the $\mhsm\lsim 200\gev$
mass region when the top quark was not thought to be so heavy.)
Second, we observe that the $VV\rta \hsm$ ($V= \wpm,Z$)
fusion cross section is at best about
10\% of the $gg\rta \hsm$ fusion cross section, and that it will
be difficult to separate at the low event rates and Higgs boson masses
relevant for a 4 TeV Tevatron.  Thus, we do not consider $VV$
fusion in our analysis. Finally, we note that the $t\anti t \hsm$ and
$b\anti b\hsm$ associated production processes are substantially
smaller (by at least factors of order 20 and 5, respectively)
than $W\hsm$ associated production, due to the relatively small $gg$
luminosity at the UT. Consequently, these processes do not appear
in the figure and are not considered further.

\vskip .15in
\noindent{\bf 2. Feasibility for SM Higgs Boson Searches
at an Upgraded Tevatron}
\vskip .05in

\noindent\undertext{A. Detector Characteristics and Acceptance Cuts}
\bigskip

It is hardly necessary to dwell on the reasons
for considering the various production/decay modes for $\hsm$
detection that we have outlined in the introduction.
Thus, we shall proceed quickly to graphs of signal and background event rates,
once some detector issues and choices have been discussed.
Our approach with regard to the detector is to consider
generic resolutions for: i) a detector very much like the
current CDF and D0 detectors; and ii) a much more optimized detector,
significantly upgraded from the current CDF and D0 detector
characteristics. For case i) we adopt energy resolutions given by:
$${\Delta E\over E}=\cases {{0.3\over\sqrt E}\oplus 0.01&  for $l=e,\mu$ \cr
                            {0.8\over\sqrt E}\oplus 0.05&  for jets$\,,$\cr}
\eqn\deltaepessimistic$$
whereas for the optimized detector, case ii), our resolutions are assumed to be
$${\Delta E\over E}=\cases {{0.2\over\sqrt E}\oplus 0.01&  for $l=e,\mu$ \cr
                            {0.5\over\sqrt E}\oplus 0.03&  for jets$\,.$\cr}
\eqn\deltaeoptimistic$$
While these latter performance
characteristics exceed those of the existing CDF and D0 detectors,
they are not dramatically better and could be achieved
in an upgraded detector for the UT.

We will also consider two possibilities for acceptance cuts.
In the first, labelled as case a), we impose the following cuts on
leptons included in our triggering or mass distributions:
$$p_T^l>20\gev;\quad |y_l|<2.5; \quad \Delta R_l>0.7; \quad \ptmiss>20\gev\,.
\eqn\casea$$
Here, $\Delta R_l$ is the minimum separation of the lepton in question
from all other jets and leptons.
The $\ptmiss$ cut is, of course, only relevant for processes with $W\rta l\nu$.
These conservative acceptance cuts will be contrasted with more
optimistic choices, labelled as case b):
$$p_T^l>10\gev;\quad |y_l|<2.5; \quad \Delta R_l>0.3; \quad \ptmiss>15\gev\,.
\eqn\caseb$$
Non-$b$ jets that are specifically utilized are required to
have:
$$p_T^j\geq15\gev;\quad |y_j|\leq 2.5;\quad \Delta R_j\geq 0.7\,,\eqn\jets$$
where $\Delta R_j$ is the separation from other jets.
We assume that any given $b$-jet can be  tagged with 30\% efficiency
and 99\% purity for
$$p_T^b\geq 15\gev;\quad |y_b|<2\,,\eqn\btagcuts$$
provided the $b$ is separated by an appropriate $\Delta R_b$ from neighboring
$b$ and light quark/gluon jets. For case a) cuts, we require
$\Delta R_b>0.7$, while for case b) we require $\Delta R_b>0.5$.
We note that trigger rate should not be a problem for the less
severe lepton acceptances.
The issue is whether or not they can be employed in the analysis for
a given type of signal without increasing the contamination from
background processes other than those we explicitly compute.
Clearly, this is at least partly a machine and detector dependent issue.

For some of the processes under consideration, the $\hsm$ signal
is to be revealed as a mass peak in the $b\anti b$ decay channel.
It is clearly important to consider the effect of the semi-leptonic
$b$ decays upon the mass resolution that can be achieved in this
channel.  In our analysis $b$ quarks are allowed to decay
semi-leptonically to $cl\nu$ according to the measured branching ratio.
The result is a moderate broadening of the $b\anti b$ mass
distribution compared to that obtained if semi-leptonic decays are
not included.  The broadening we compute could be either greater
or less than that which will in fact occur.  The fact that
the $b$-jet puts only a fraction of its momentum into
the $B$-meson that actually decays, means that the neutrinos
from the primary $b\rta cl\nu$ decay of our computation are
on the average more energetic than would be predicted in a full
simulation.  This effect means our broadening could be an overestimate.
On the other hand, we do not include $c$ decays, which sometimes
also yield neutrinos (although relatively soft ones).  Such neutrinos
would cause additional broadening.  We believe that our approximate treatment
is adequate for this first survey of Higgs physics at an upgraded
Tevatron. Of course, one might also wish to consider the possibility
of eliminating semi-leptonic $b$-decays by rejecting events in
which leptons are visible within the tagged $b$-jet(s).
This will yield a narrower $b\anti b$ distribution, but some events
will be lost from the signal and background.  Our estimate is
that this would not result in a significant improvement in the observability
of the signals being considered.

\bigskip
\noindent\undertext{B. Detailed Results}
\bigskip

\FIG\whbb{
The $M_{b\anti b}$ event rate distribution for
$\wpm\hsm\rta l\nu b\anti b$ associated production is plotted
in 5 GeV bins, for the indicated lepton and $b$-jet cuts (see text
for more details). Signals for $\mhsm=60$, 80, 100, 110, 120, 130
and $150\gev$ are shown as the solid histograms.
Also shown are the electroweak $WZ\rta l \nu b\anti b$
background (dots) and the mixed QCD/EW continuum backgrounds from
$W+2j$ (dash-dot) and $W+b\anti b$ (dashes) production. Semi-leptonic
$b$-decays are included. $b$-tagging is assumed to have 30\% efficiency
and 99\% purity for the stated cuts. Graphs for both 1 and 2 $b$-tags
are shown. A `K' factor of 1.2 is included for the $\hsm$ signals.
Conservative detector/cuts case i)-a) is employed.
}
\topinsert
\vbox{\phantom{0}\vskip 4in
\phantom{0}
\vskip .5in
\hskip -0pt
\special{ insert scr:whbb_1btag_4tev_ppbar_willen.ps}
\vskip 3.6in
\hskip 5pt
\special{ insert scr:whbb_2btag_4tev_ppbar_willen.ps}
\vskip -1.45in }
\centerline{\vbox{\hsize=12.4cm
\Tenpoint
\baselineskip=12pt
\noindent
Figure~\whbb:
The $M_{b\anti b}$ event rate distribution for
$\wpm\hsm\rta l\nu b\anti b$ associated production is plotted
in 5 GeV bins, for the indicated lepton and $b$-jet cuts (see text
for more details). Signals for $\mhsm=60$, 80, 100, 110, 120, 130
and $150\gev$ are shown as the solid histograms.
Also shown are the electroweak $WZ\rta l \nu b\anti b$
background (dots) and the mixed QCD/EW continuum backgrounds from
$W+2j$ (dash-dot) and $W+b\anti b$ (dashes) production. Semi-leptonic
$b$-decays are included. $b$-tagging is assumed to have 30\% efficiency
and 99\% purity for the stated cuts. Graphs for both 1 and 2 $b$-tags
are shown. A `K' factor of 1.2 is included for the $\hsm$ signals.
Conservative detector/cuts case i)-a) is employed.
}}
\endinsert

\noindent\undertext{$\wpm\hsm\rta l\nu b\anti b$}
\vskip .05in

The most promising channel for $\hsm$ detection in the mass region
$\mhsm\lsim 120\gev$ is $\wpm\hsm\rta l\nu b\anti b$
associated production, leading to the $lb\anti b X$ final
state. In Fig.~\whbb\ we plot the event rate
distribution as a function of the $b\anti b$ mass, in the case of both
1-$b$ and 2-$b$ tagging. For these plots we have demanded that the $l$
satisfy the conservative acceptance cuts a) outlined earlier.
Resolutions employed are those for the conservative detector i).
Semi-leptonic $b$-decays are incorporated using the procedure described
above.  (The effect of not including them is to narrow the signal
peak, increasing the peak height by about 10\%.)
A QCD `K' factor of 1.2 has been included in the signal rates; no K
factors are included in the background rates.

 \TABLE\whbbtablei{}

 \midinsert
 \titlestyle{\twelvepoint
 Table \whbbtablei: We tabulate $S$, $B$, and $S/\sqrt B$
for $W\hsm\rta lb\anti b X$,
at $p\anti p$ and $pp$ colliders for $L=30\fbi$ and $\sqrt s=4\tev$, for
a series of $\mhsm$ values, using the tabulated mass bins.
Also given is the $L$ (in $\fbi$) required for $S/\sqrt B=5$.  This table
is for a conservative detector, case i), and conservative cuts a).
}
 \bigskip
\def\tstrut{\vrule height 14pt depth 4pt width 0pt}
 \thicksize=0pt
 \hrule \vskip .04in \hrule
 \begintable
 \ | \ | \ | \multispan{4} \tstrut\hfil $p\anti p$ \hfil |
        \multispan{4} \tstrut\hfil $pp$ \hfil \nr
 $\mhsm$  | $\Delta M$ | $n_{btag}$ | $S$ & $B$ & $S/\sqrt B$ & $L(5\sigma)$ |
                         $S$ & $B$ & $S/\sqrt B$ & $L(5\sigma)$ \cr
 60 | 20 | 1 | 2300 & 30358 & 13.2 & 4.3 | 1731 & 26663 & 10.6 & 6.7 \nr
 60 | 20 | 2 | 406 & 780 & 14.5 & 3.6 | 306 & 594 & 12.5 & 4.8 \cr
 80 | 20 | 1 | 1200 & 20844 & 8.3  & 10.8 | 850  & 18020 & 6.3  & 18.7 \nr
 80 | 20 | 2 | 212 & 480 & 9.7  & 8.0 | 150 & 343 & 8.1  & 11.4 \cr
 100 | 30 | 1 | 755  & 19996 & 5.3  & 26.3 | 503  & 17747 & 3.8  & 52.6 \nr
 100 | 30 | 2 | 133 & 425 & 6.5  & 18.0 | 89   & 306 & 5.1  & 29.1 \cr
 110 | 30 | 1 | 549  & 15931 & 4.3  & 39.7 | 354  & 14238 & 3.0  & 85.1 \nr
 110 | 30 | 2 | 97  & 330 & 5.3  & 26.4 | 63   & 232 & 4.1  & 44.5 \cr
 120 | 30 | 1 | 385  & 12985 & 3.4  & 65.9 | 241  & 11650 & 2.2  & 151  \nr
 120 | 30 | 2 | 68  & 260 & 4.2  & 42.3 | 43   & 179 & 3.2  & 74.2 \cr
 130 | 30 | 1 | 244  & 10460 & 2.4  & 148  | 148  &  9358 & 1.5  & 321  \nr
 130 | 30 | 2 | 43  & 207 & 3.0  & 84.2 | 26   & 141 & 2.2  & 156
\endtable
 \hrule \vskip .04in \hrule
 \endinsert

 \TABLE\whbbtableii{}

 \midinsert
 \titlestyle{\twelvepoint
 Table \whbbtableii:
We tabulate $S$, $B$, and $S/\sqrt B$ for $W\hsm\rta lb\anti b X$,
at $p\anti p$ and $pp$ colliders for $L=30\fbi$ and $\sqrt s=4\tev$, for
a series of $\mhsm$ values, using the tabulated mass bins.
Also given is the $L$ (in $\fbi$) required for $S/\sqrt B=5$. This table
is for an optimistic detector, case ii), and optimistic acceptance cuts b).
}
 \bigskip
\def\tstrut{\vrule height 14pt depth 4pt width 0pt}
 \thicksize=0pt
 \hrule \vskip .04in \hrule
 \begintable
 \ | \ | \ | \multispan{4} \tstrut\hfil $p\anti p$ \hfil |
        \multispan{4} \tstrut\hfil $pp$ \hfil \nr
 $\mhsm$  | $\Delta M$ | $n_{btag}$ | $S$ & $B$ & $S/\sqrt B$ & $L(5\sigma)$ |
                         $S$ & $B$ & $S/\sqrt B$ & $L(5\sigma)$ \cr
 60 | 10 | 1 | 2649 & 19593 & 18.9 & 2.1 | 1994 & 17777 & 15.0 & 6.7 \nr
 60 | 10 | 2 | 467 & 505 & 20.8 & 1.7 | 352 & 375 & 18.2 & 2.3 \cr
 80 | 10 | 1 | 1300 & 14854 & 10.7 & 6.6  | 924  & 12005 & 8.4  & 10.6 \nr
 80 | 10 | 2 | 229 & 320 & 12.8 & 4.6 | 163 & 228 & 10.8 & 6.4  \cr
 100 | 20 | 1 | 948  & 17664 & 7.1  & 14.7 | 634  & 15602 & 5.1  & 28.9 \nr
 100 | 20 | 2 | 167 & 372 & 8.7  & 10.0 | 112  & 263 & 6.9  & 15.6 \cr
 110 | 20 | 1 | 691  & 14270 & 5.8  & 22.4 | 451  & 12610 & 4.0  & 46.5 \nr
 110 | 20 | 2 | 122 & 278 & 7.3  & 14.0 | 80   & 196 & 5.7  & 23.2 \cr
 120 | 20 | 1 | 483  & 11373 & 4.5  & 36.6 | 306  & 10118 & 3.0  & 81   \nr
 120 | 20 | 2 | 85  & 217 & 5.8  & 22.4 | 54   & 155 & 4.3  & 40.0 \cr
 130 | 20 | 1 | 309  & 9470  & 3.2  & 74.4 | 190  &  8020 & 2.1  & 167  \nr
 130 | 20 | 2 | 55  & 182 & 4.0  & 46.1 | 34   & 127 & 3.0  & 84.5
\endtable
 \hrule \vskip .04in \hrule
 \endinsert

In order to assess the observability of these signals, we have
computed signal and background rates
by simply looking for a peaking in $M_{b\anti b}$ and subtracting
a smooth background estimated from surrounding mass bins.
The statistical significances that could be achieved
after optimizing the signal mass intervals
appear in Tables~\whbbtablei\ and \whbbtableii.
The tables give signal and background rates, $S$ and $B$,
along with $S/\sqrt B$ for $L=30\fbi$ at both a $p\anti p$ and
a $pp$ collider.  Also shown is the $L$ required to achieve
$S/\sqrt B=5$.  Results are presented for both single and double
$b$-tagging with 30\% efficiency and 99\% purity.
The numbers in Table~\whbbtablei\ are for
the conservative detector case i), and employ the conservative set
of acceptance cuts a).  Table~\whbbtableii\ is for the more optimized
detector case ii) and more optimistic cuts b).

Overall, the tables show that the $Wb\anti b$ signal
is almost certainly viable for $\mhsm\lsim 100\gev$, \ie\ over a range
comparable to that for which the $\hsm$ will be found at LEP-200.
The most difficult signal
in this general mass region would be for $\mhsm\sim \mz$. As indicated
by the dotted histogram in Fig.~\whbb, the $WZ$ and $W\hsm$ final states yield
mass peaks of very similar magnitude.  Thus, the $\hsm$ would
have to be recognized as an excess over that which would be expected
from the $WZ$ continuum.  We do not believe that this would be especially
difficult, given that the $WZ$ continuum can be normalized by
using other channels, in particular the three-lepton channel.

The most crucial question is whether
or not a 4 TeV Tevatron can go beyond the reach of LEP-200.
We see from the tables that to reach $\mhsm=110\gev$,
either $L>10\fbi$ is required or the optimized detector ii)
and less stringent acceptance cuts b) must prove possible.
Since a comparison of the single $b$-tag and double $b$-tag results in the
two tables makes clear that double-tagging has a clear advantage,
\foot{This advantage would be even greater if the purity of $b$-tagging
were not as great as assumed.  This would be especially likely
at instantaneous luminosities substantially above
${\cal L}=10^{33}~cm^{-2}\,s^{-1}$, such as clearly required
for $\mhsm\gsim 110\gev$.}
let us quote the corresponding numbers. To obtain a $\nsd=5$ signal
at $\mhsm=110\gev$ requires $26\fbi$ in $p\anti p$ collisions and $44\fbi$
in $pp$ collisions, for the conservative detector/cuts choices,
Table~\whbbtablei.
The former is feasible after three or so years of running, while
the latter would require an instantaneous luminosity ${\cal L}$
approaching the ${\cal L}=10^{34}~cm^{-2}\,s^{-1}$ level for which
a redesigned $pp$ collider might be built.  To maintain purity of
$b$-tagging, operation somewhat below this level would be desirable
and, as we see, adequate. To detect a signal for $\mhsm=120\gev$
will be exceedingly difficult without an optimized detector and an
ability to employ weaker cuts.  From Table~\whbbtableii\
we see that an $\mhsm=120\gev$ signal at the $\nsd=5$ level
in these optimal circumstances requires $L=22\fbi$ and $40\fbi$
for $p\anti p$ and $pp$ collisions, respectively, \ie\ very similar
to the requirements for $\mhsm=110\gev$ with conservative detector/cuts
choices.

Thus, we conclude that $\mhsm\lsim 110\gev$ could
be probed in the $W\hsm\rta l b\anti b X$ mode, reaching
possibly as high as $\mhsm=120\gev$ if an optimized detector
is available and optimal acceptance cuts can be employed.

\vskip .05in
\noindent\undertext{$Z\hsm\rta 2l (2\nu) b\anti b$}
\vskip .05in

The analogous signal in the $Z\hsm\rta 2lb\anti b$ channel
is also possibly interesting.  The $M_{b\anti b}$ distributions
for signal and backgrounds are qualitatively similar to those appearing
in Fig.~\whbb, but the overall event rates are much lower --- roughly
by a factor of 9--10 in the case of the signal.  As a result, even
at $\mhsm=60\gev$, for optimized detector case ii) and cuts b)
only $\nsd\sim 6$ is achieved in both the 1 and 2 $b$-tag
cases.  By $\mhsm=100\gev$, $\nsd$ has declined to $\sim 2.5$.

\FIG\fourl{
The $gg\rta \hsm\rta ZZ^{(*)}\rta 4l$ ($l=e,\mu$) signal and
$q\anti q \rta ZZ \rta 4l$ background as a function of the four-lepton mass,
$M_{4l}$, in 5 GeV bins.  Higgs boson signals for $\mhsm=130$, 150, 170, 200,
230, 270, 300 and $400\gev$ are illustrated, after
including a `K' factor of 1.5.
Effects of the Higgs boson width are incorporated in order to obtain the
correct signal shapes at large $\mhsm$ values.
Optimistic lepton resolution/acceptance choices, case ii)-b), are employed.
}
\midinsert
\vbox{\phantom{0}\vskip 5.0in
\phantom{0}
\vskip .5in
\hskip -0pt
\special{ insert scr:4l_4tev.ps}
\vskip -1.45in }
\centerline{\vbox{\hsize=12.4cm
\Tenpoint
\baselineskip=12pt
\noindent
Figure~\fourl:
The $gg\rta \hsm\rta ZZ^{(*)}\rta 4l$ ($l=e,\mu$) signal and
$q\anti q \rta ZZ \rta 4l$ background as a function of the four-lepton mass,
$M_{4l}$, in 5 GeV bins.  Higgs boson signals for $\mhsm=130$, 150, 170, 200,
230, 270, 300 and $400\gev$ are illustrated, after
including a `K' factor of 1.5.
Effects of the Higgs boson width are incorporated in order to obtain the
correct signal shapes at large $\mhsm$ values.
Optimistic lepton resolution/acceptance choices, case ii)-b), are employed.
}}
\endinsert

One could also consider the $Z\hsm\rta 2\nu b\anti b$ channel.
This channel has a good event rate, but is subject to many detector-dependent
backgrounds and to triggering problems.
One background is $gb\anti b$ production,
where the $g$ disappears down the beam-pipe hole or fragments
to very soft particles that are not reconstructed as jets.  Since
the signal for a light $\hsm$ is concentrated at low $Zb\anti b$
subprocess energies, it would be necessary to retain events
for which the missing energy lies significantly lower than
$\mz$. But, it is far from
clear that the very high $gb\anti b$ rate can be sufficiently reduced
unless the lower threshold for the missing energy is rather
substantial.  Indeed,
missing energy would provide one of the main triggers for this mode,
and it is not clear how far below $\mz$ the threshold can be
set while providing even an acceptable trigger rate. This is a
very detector-dependent issue that we have not pursued further.
We are very doubtful that this mode can be useful, and it surely
would never be competitive with the $Wb\anti b$ mode that
we have analyzed in detail.

 \TABLE\fourltable{}
 \topinsert
 \titlestyle{\twelvepoint
 Table~\fourltable: For $ZZ\rta 4l$,
we tabulate as a function of Higgs boson mass: the signal and background
rates $S$ and $B$ (summed over a 10 GeV interval), for $p\anti p$ ($pp$),
for $L=30\fbi$; the associated $S/\sqrt B$ values;
and the $L$ (in $\fbi$) required for a $S/\sqrt B=5$ signal level.
The background rate for $M_{4l} < 2m_Z^{}$ is negligible.
Optimistic lepton resolution/acceptance choices, case ii)-b), are employed.
}
 \bigskip
\def\tstrut{\vrule height 14pt depth 4pt width 0pt}
 \thicksize=0pt
 \hrule \vskip .04in \hrule
 \begintable
 $\mhsm$ | $S$ | $B$ | $S/\sqrt B$ | $L(5\sigma)$ \cr
 130     | 5   |  -  |  -          | -            \cr
 150     | 13  |  -  |  -          | -            \cr
 170     | 4   |  -  |  -          | -            \cr
 200     | 28  | 28(21) | 5.3(6.2) | 27(20)       \cr
 230     | 20  | 21(15) | 4.3(5.2) | 41(28)       \cr
 270     | 12  | 12(8)  | 3.5(4.3) | 61(41)       \cr
 300     | 8   | 8(5)   | 2.8(3.6) | 96(58)
\endtable
 \hrule \vskip .04in \hrule
 \endinsert

\vskip .05in
\noindent\undertext{$gg\rta\hsm\rta ZZ^*\rta 4l$}
\vskip .05in

The next channel we consider is $gg\rta\hsm\rta ZZ^*\rta 4l$ for $\mhsm<2\mz$.
Here, it is absolutely critical to accept leptons with
as low a momentum as possible, since the light Higgs boson signals generally
yield leptons that are not terribly energetic.  Thus, we employ the
case b) cuts delineated in Eq.~\caseb. The acceptance cuts of case a) are
not considered since too much of the light Higgs boson signals of interest
would be eliminated for this mode to have even a chance of yielding
a signal. As noted earlier, for $\mhsm<2\mz$ there is no significant background
so long as the two leptons that do not reconstruct to an on-shell
$Z$ are constrained to have significant mass.
\refmark{\gkw,\sdctdr,\cmsatlas}
The signal automatically leads to large $2l$ mass values for the leptons from
the $Z^*$, whereas backgrounds from virtual photons yield very low $2l$
mass values. Meanwhile, the continuum $ZZ^*$ background is negligible.
The signal $4l$ event rate as a function
of the $4l$ mass is plotted in 5 GeV
bins in Fig.~\fourl, for several $\mhsm$ values.  The signal
rates include a `K' factor of 1.5.
Lepton momenta have been smeared using
the more optimistic resolution values of detector case ii), and event numbers
reflect the optimistic acceptance cuts b).
In Table~\fourltable\ we give event rates after summing over a 10 GeV
mass interval centered on $\mhsm$.
We see immediately that, even for these optimistic choices,
there are very few events.
For the best case of $\mhsm\sim 150\gev$, there are only about
13 events altogether for $L=30\fbi$.
Increasing the threshold for lepton detection to the more conservative
case a) value would further decrease the event rates.
Finally, including a charged lepton tracking efficiency
(for tracks within the already imposed fiducial cuts) of order 0.95
for each lepton would decrease the rates by a factor of $\sim 0.8$.
Thus, the feasibility of detecting the $\hsm$ in this very clean
channel is clearly limited by the small event rate.

For $\mhsm>2\mz$, the $\hsm \rta ZZ\rta 4l$ event rate
increases somewhat, but the $ZZ\rta 4l$ continuum background enters,
as shown in Fig.~\fourl\ and tabulated in Table~\fourltable.
For the optimum $\mhsm=200\gev$ choice
and optimistic lepton resolution and acceptance cuts,
$S/\sqrt B = 28/{\sqrt 28}\sim 5$ is achieved for $L=30\fbi$
(keeping the two central 5 GeV bins). After including
our estimate of 0.8 for the net efficiency of
finding all four lepton tracks, this signal retreats below the $5\sigma$ level.
However, the cleanliness of this channel is such that the signal
still could probably be observed even for peaks down to the $3\sigma$ level,
provided there are an adequate number of events,
say, at least 15 or so, in the signal peak.
With this more optimistic criteria, even after including the 0.8
track-finding efficiency the $\hsm$ would be detectable for
$2 m_Z^{} \lsim\mhsm \lsim 250\gev$.
However, changing the lepton acceptance cuts to the more
conservative case a) value would significantly decrease the event rates to an
extent that it would be difficult to detect the $\hsm$ for any value
of $\mhsm$. For instance, for $200\lsim\mhsm\lsim 270\gev$,
the change from case b) to case a) cuts results in a decrease
of event rate by a factor of about 2.
What would it take to reach $\mhsm=300\gev$ in this mode?
Certainly, the situation is best for a $pp$ collider,
which, in any case, is the only type of collider
that could reach the required luminosities.  Table~\fourltable\
shows that nominal statistical significance (before
including tracking efficiencies)
of $S/\sqrt B \sim 5$ could be achieved for $L\sim 60\fbi$
in $pp$ collisions, {\it provided} that optimistic acceptance cuts
and both $e$ and $\mu$ signals could be employed at the
high instantaneous luminosities required.

\FIG\ljj{
The $gg\rta \hsm\rta WW \rta l\nu jj$ signal and various
background event rates as a function of the
cluster transverse mass, $M_{ljj}^{clus}$, in 5 GeV bins, for $L=30\fbi$,
$\sqrt s=4\tev$ $p\anti p$ collisions. Signals for
$\mhsm=130$, $150$, $170$, $200$, $230$, $270$, and $300\gev$ are shown.
The $t\anti t$ background is shown before vetoing against additional
central jets ($\mt=170\gev$). A QCD `K' factor of 1.5 is included
in the signal rates. Optimistic detector resolutions and acceptance
cuts are employed.
}
\midinsert
\vbox{\phantom{0}\vskip 5.0in
\phantom{0}
\vskip .5in
\hskip -0pt
\special{ insert scr:ljj_4tev_ppbar.ps}
\vskip -1.45in }
\centerline{\vbox{\hsize=12.4cm
\Tenpoint
\baselineskip=12pt
\noindent
Figure~\ljj:
The $gg\rta \hsm\rta WW \rta l\nu jj$ signal and various
background event rates as a function of the
cluster transverse mass, $M_{ljj}^{clus}$, in 5 GeV bins, for $L=30\fbi$,
$\sqrt s=4\tev$ $p\anti p$ collisions. Signals for
$\mhsm=130$, $150$, $170$, $200$, $230$, $270$, and $300\gev$ are shown.
The $t\anti t$ background is shown before vetoing against additional
central jets ($\mt=170\gev$). A QCD `K' factor of 1.5 is included
in the signal rates. Optimistic detector resolutions and acceptance
cuts are employed.
}}
\endinsert

\vskip .05in
\noindent\undertext{$gg \rta \hsm\rta WW\rta l\nu jj$}
\vskip .05in

Next, we discuss the $\hsm\rta WW\rta l\nu jj$
channel.  Aside from the irreducible $WW$ continuum background,
there are also backgrounds from the mixed QCD/EW $Wjj$ channels
and from $t\anti t\rta WWb\anti b$ production.  In order to reduce
these backgrounds it is necessary to impose a cut on the $jj$
mass in the vicinity of $\mw$.  We have accepted events which
(after smearing) yield $jj$ mass between $\mw\pm 10\gev$.
(For our resolutions, this choice is near optimum.)  The resulting
signal and event rates are plotted in Fig.~\ljj\ as a function
of the cluster transverse mass defined by:
$M_{ljj}^{cluster}\equiv \sqrt{m_{ljj}^2+p_{T\,ljj}^2}
+\ptmiss$.  This figure is for the optimistic detector/cuts scenario
ii)-b).

Mass peaks with large numbers of events
emerge for the signal, but background event rates, especially from
the mixed QCD/EW $Wjj$ process, are very much larger.  The $t\anti t$
background can, however, be effectively eliminated by vetoing events
with extra jets in the central region (at least one of
the $b$'s from the $t$ decays nearly always will appear as an energetic
central jet). In addition, we have found several cuts that
help to increase $S/B$ and the statistical significance of the
mass peaks.  First, consider the ratio of the total 3-momentum of
the less energetic jet relative to that of the more energetic jet,
$r\equiv|\vec p_j^{\,min}|/|\vec p_j^{\,max}|$.  A cut of $r\geq 0.3$
reduces the $Wjj$ background by about 10\% without affecting the signal
rates significantly.  Second, consider the $\Delta R$ separation
between the two jets.  Higgs boson signals always fall in a well defined
range of $\Delta R$, whereas the $Wjj$ and $WW$ backgrounds
have a larger spread, even if one retains only $M_{ljj}^{cluster}$ values
in the vicinity of $\mhsm$.  Thus, we impose a $\Delta R$ cut that depends
upon the Higgs boson mass.  (Practically, the experimental groups would
examine the $M_{ljj}^{cluster}$ distribution for each of the proposed
cuts and look for a peak in the corresponding mass region.)
The best choices are
$\Delta R\in [2.8,3.5]$, $[2.6,3.5]$, $[2.3,3.5]$, $[2.0,3.5]$,
$[2.0,3.5]$, $[1.8,3.5]$, $[1.5,3.0]$, $[1.2,3.0]$, $[1.0,3.0]$,
for $\mhsm=110$, 120, 130, 150, 170, 200, 230, 270 and $300\gev$,
respectively.

Finally, we have examined the distribution for
$\cos\phi_{min}$, where $\phi_{min}$ is the smaller of the transverse plane
azimuthal angles between the observed lepton and the two-jets.
(Note that we cannot determine which jet is the fermion vs. anti-fermion,
so that the analogue of $\phi_{2l}$ that will be
employed in the $2l$ analysis is unfortunately not available.)
At the higher masses of $\mhsm=270$ and $300\gev$, we find that
the signal distributions in $\cos\phi_{min}$ develop a double peaked
structure, with a $\cos\phi_{min}$ peak below 0.5 as well as the
peak near 1 that is present at all masses.  Meanwhile, the backgrounds
only exhibit a peak above 0.5, and are quite suppressed for
$\cos\phi_{min}<0.5$.  Thus, it is highly advantageous to impose
a cut of $\cos\phi_{min}<0.5$ in searching for a Higgs boson
with $\mhsm\gsim 270\gev$.

 \TABLE\ljjtablei{}

 \midinsert
 \titlestyle{\twelvepoint
 Table \ljjtablei: For $WW\rta l\nu jj$,
we tabulate $S$, $B$, and $S/\sqrt B$ for $L=30\fbi$,
for $p\anti p$ and $pp$ colliders at $\sqrt s=4\tev$, for
a series of $\mhsm$ values, using the tabulated mass bins.
Also given is the $L$ (in $\fbi$) required for $S/\sqrt B=5$.  This table
is for the conservative detector, case i), and conservative
acceptance cuts a).
}
 \bigskip
\def\tstrut{\vrule height 14pt depth 4pt width 0pt}
 \thicksize=0pt
 \hrule \vskip .04in \hrule
 \begintable
 \ | \ | \ | \multispan{3} \tstrut\hfil $p\anti p$ \hfil |
             \multispan{3} \tstrut\hfil $pp$ \hfil \nr
 $\mhsm$  | $\Delta M$ | $S$ | $B$ & $S/\sqrt B$ & $L(5\sigma)$ |
                               $B$ & $S/\sqrt B$ & $L(5\sigma)$ \cr
 130 | 20 | 271  | 16200  & 2.1  & 165  | 12800  & 2.4  & 131  \cr
 150 | 30 | 1539 | 133000 & 4.2  & 42   | 109000 & 4.7  & 35   \cr
 170 | 30 | 4542 | 251000 & 9.1  & 9.1  | 226000 & 9.5  & 8.2  \cr
 200 | 30 | 1918 | 200000 & 4.3  & 41   | 174000 & 4.6  & 35   \cr
 230 | 40 | 1290 | 113000 & 3.8  & 51   | 102000 & 4.0  & 46   \cr
 270 | 40 | 724  | 42300  & 3.5  & 61   | 37600  & 3.7  & 54   \cr
 300 | 40 | 558  | 20300  & 3.9  & 49   | 20700  & 3.9  & 50
\endtable
 \hrule \vskip .04in \hrule
 \endinsert

 \TABLE\ljjtableii{}

 \midinsert
 \titlestyle{\twelvepoint
 Table \ljjtableii: For $WW\rta l\nu jj$,
we tabulate $S$, $B$, and $S/\sqrt B$ for $L=30\fbi$,
for $p\anti p$ and $pp$ colliders at $\sqrt s=4\tev$, for
a series of $\mhsm$ values, using the tabulated mass bins.
Also given is the $L$ (in $\fbi$) required for $S/\sqrt B=5$.  This table
is for the optimized detector, case ii), and optimistic
acceptance cuts b).
}
 \bigskip
\def\tstrut{\vrule height 14pt depth 4pt width 0pt}
 \thicksize=0pt
 \hrule \vskip .04in \hrule
 \begintable
 \ | \ | \ | \multispan{3} \tstrut\hfil $p\anti p$ \hfil |
             \multispan{3} \tstrut\hfil $pp$ \hfil \nr
 $\mhsm$  | $\Delta M$ | $S$ | $B$ & $S/\sqrt B$ & $L(5\sigma)$ |
                               $B$ & $S/\sqrt B$ & $L(5\sigma)$ \cr
 130 | 20 | 834  | 47500  & 3.8  & 51   | 30664  & 4.8  & 33   \cr
 150 | 30 | 2633 | 215600 & 5.7  & 23   | 169300 & 6.4  & 18   \cr
 170 | 30 | 6736 | 331200 & 11.7 & 5.5  | 296600 & 12.4 & 4.9  \cr
 200 | 30 | 2767 | 239000 & 5.7  & 23   | 225000 & 5.8  & 22   \cr
 230 | 40 | 1739 | 144000 & 4.6  & 36   | 125000 & 4.9  & 31   \cr
 270 | 40 | 887  | 46500  & 4.1  & 44   | 57200  & 3.7  & 55   \cr
 300 | 40 | 697  | 22100  & 4.1  & 44   | 31400  & 3.9  & 49
\endtable
 \hrule \vskip .04in \hrule
 \endinsert

After imposing these ($\mhsm$-dependent) cuts
(which typically enhance the signal to background
ratio by about a factor of 2), the signal and background
rates, and nominal statistical significances
($\nsd=S/\sqrt B$) at $L=30\fbi$, are
tabulated in Tables~\ljjtablei\ and \ljjtableii,
for conservative detector/cuts i)-a) and optimistic
detector/cuts ii)-b), respectively.
Also given are the $L$ values required
for a $\nsd=5$ level signal.  Results for both $p\anti p$ and $pp$
collisions are tabulated. The $\nsd$ values are, of course, computed purely
on a statistical basis.  We notice that $S/B$ ratios are typically
at the 1\% level, so that systematics will play a crucial
role.  Even though there are distinct mass peaks (in contrast
to the shapeless distributions we shall
encounter in the $WW\rta 2l2\nu$ channel)
the very small $S/B$ level will mean that
the shape of the background distribution must be very well understood.
Uncertainties in the theoretical computations and detector
response and efficiencies are likely to be large enough
that extraction of these $ljj$ signals may be very difficult, especially
in the lower mass region $\mhsm\lsim 170$ where the background
does not have a simple shape, and depends significantly on
cut thresholds \etc\

\FIG\twol{
The $gg\rta \hsm\rta WW^{(*)}\rta 2l 2\nu$ signal and
$q\anti q \rta WW \rta 2l 2\nu$ background event rates as a function of the
two-lepton mass, $M_{2l}$, in 5 GeV bins.  Signals for
$\mhsm=120$, $130$, $150$, $170$, and $200\gev$ are shown.
A $\cos\phi_{2l}>0$ cut is imposed to eliminate the background from $\tauptaum$
continuum pair production. QCD `K' factors of 1.5 and 1.1, respectively, are
included in the $gg$ fusion and $WW$ continuum event rates.
}
\midinsert
\vbox{\phantom{0}\vskip 5.0in
\phantom{0}
\vskip .5in
\hskip -0pt
\special{ insert scr:2l_4tev_strcuts.ps}
\vskip -1.45in }
\centerline{\vbox{\hsize=12.4cm
\Tenpoint
\baselineskip=12pt
\noindent
Figure~\twol:
The $gg\rta \hsm\rta WW^{(*)}\rta 2l 2\nu$ signal and
$q\anti q \rta WW \rta 2l 2\nu$ background event rates as a function of the
two-lepton mass, $M_{2l}$, in 5 GeV bins.  Signals for
$\mhsm=120$, $130$, $150$, $170$, and $200\gev$ are shown.
A $\cos\phi_{2l}>0$ cut is imposed to eliminate the background from $\tauptaum$
continuum pair production. QCD `K' factors of 1.5 and 1.1, respectively, are
included in the $gg$ fusion and $WW$ continuum event rates.
}}
\endinsert

{}From Tables~\ljjtablei\ and \ljjtableii\ it seems that for $\mhsm$ between
about $150\gev$ and $200\gev$ there would be some possibility
of discovery in this mode at either a $p\anti p$ or $pp$ collider,
with the optimistic detector/cuts choices yielding a substantially
better chance. However, even the $\mhsm=170\gev$ signal does not
reach a level that one can say would certainly be seen, given
the above-discussed systematic uncertainties (that
are not reflected in the nominal $S/\sqrt B$ values quoted).
And, going beyond the $150-200\gev$
mass interval would surely be extremely difficult given the small $S/B$
ratios. In particular, even the nominal $S/{\sqrt B}$ values for
masses $230\gev$ and above show a slow deterioration in the likelihood
for discovery.  Keeping in mind that the signal is becoming quite
broad in this region, so that systematics would play
a very major role, it is not reasonable to suppose that
$\mhsm$ much above $200\gev$ could be detected.
Of course, we cannot rule out the possibility that there are other cuts
which would improve the situation.

\vskip .05in
\noindent\undertext{$gg\rta\hsm\rta WW\rta 2l 2\nu$}
\vskip .05in

A less promising, but not necessarily useless,
channel for $\hsm$ detection is $gg\rta\hsm\rta
WW^*\rta 2l 2\nu$.  As for the $ZZ\rta 4l$ signal, it is critical
to accept low-momentum leptons, so we employ case b) acceptance
cuts. We also implicitly assume that only events with very low jet activity
will be accepted.  This eliminates $t\anti t$ backgrounds.
The $2l$ mass distributions for a variety
of $\mhsm$ values are illustrated in Fig.~\twol, where we have
included a QCD `K' factor of 1.5 for the `0-jets' $gg$ fusion reaction.
Also shown is the $WW$ continuum contribution with a K factor of 1.1
for the 0-jets restriction.  In both cases, we have imposed a cut
on the azimuthal angle between the two leptons of $\cos\phi_{2l}>0$.
This cut has two important functions.  First, it eliminates an
otherwise very large background from $\tauptaum$ pair production
(which we computed including the $p_T$
distribution of the pair as obtained by simulating standard resummation
techniques). Second, the $WW$ continuum $\cos\phi_{2l}$ distribution
is strongly peaked for $\cos\phi_{2l}\sim -1$, whereas the light
Higgs boson signals exhibit peaking also for $\cos\phi_{2l}\sim +1$.
(For $\mhsm\gsim 230\gev$, this latter is no longer true and the cut
is better chosen nearer $-1$; \eg\ $\cos\phi_{2l}>-0.9$ eliminates
most of the $\tauptaum$ background. However, even this
bit of optimization does not raise $\mhsm\gsim 230\gev$ signals
to an observable level.)

 \TABLE\twoltable{}
 \topinsert
 \titlestyle{\twelvepoint
 Table \twoltable: For $WW\rta 2l2\nu$,
we tabulate as a function of Higgs boson mass: the optimum mass
interval for detecting an excess of events; the signal and background
rates $S$ and backgrounds $B$, for $p\anti p$ ($pp$),
for $L=30\fbi$ summed over that interval; the associated
$S/\sqrt B$ values, as the
absolute upper bound on the observability of the signal, ignoring
the systematics issues discussed in the text; and finally the $S/B$
ratios as an indicator of the level of systematics difficulty.
}
 \bigskip
\def\tstrut{\vrule height 14pt depth 4pt width 0pt}
 \thicksize=0pt
 \hrule \vskip .04in \hrule
 \begintable
 $\mhsm$ | Mass Interval | $S$ | $B$ | $S/\sqrt B$ | $S/B$ \cr
 120 | $7-43$ | 133 | 3302(2102) | 2.3(2.9) | 0.040(0.063) \cr
 130 | $12-53$ | 297 | 3956(2521) | 4.7(5.9) | 0.075(0.12) \cr
 150 | $12-68$ | 674 | 4697(2987) | 9.8(12.3) | 0.14(0.22) \cr
 170 | $12-83$ | 993 | 5150(3258) | 13.8(17.4) | 0.19(0.30) \cr
 200 | $12-113$ | 434 | 5626(3524) | 5.8(7.3) | 0.077(0.12)
\endtable
 \hrule \vskip .04in \hrule
 \endinsert

Assessing the observability of the signals illustrated in Fig.~\twol\
is difficult. Due to the broad nature
of the $M_{2l}$ signal distribution, the signals and background
are very similar in shape, and it is not
possible to simply look for a mass peak.
Thus, it is necessary to detect an event excess integrated over
a fairly broad mass interval relative to expectations in the
absence of a Higgs resonance.  Since
the signal to background ratios are not large, this will require
an extremely accurate determination of the $WW$ continuum normalization.
In order to quantify these difficulties, we present in Table~\twoltable\
the $L=30\fbi$ signal and background rates for optimally chosen
$M_{2l}$ intervals as a function of $\mhsm$.  Also given is
the nominal statistical significance that could be achieved if
there were no systematic uncertainty in the background level.
The accuracy below which the systematic uncertainty in the background
would have to be reduced in order that statistics dominate is
indicated by the $S/B$ ratio, also tabulated.

To what level can the systematic background uncertainty be reduced?
Let us assume that QCD corrections to the $WW$ continuum are computed
to 2-loops (note that we require the `0-jets'
component of the $WW$ continuum),
that precision quark and antiquark distributions
are available from HERA data, and that gluon resummation
technology continues to improve.  It is then not inconceivable
that the {\it shape} of the $M_{2l}$ distribution could be predicted
with good accuracy. However, the predicted {\it normalization} would almost
certainly have a substantial uncertainty. Thus, one would make an
experimental determination of absolute normalization by measuring the
$2l$ spectrum at large $M_{2l}$. In combination with the shape
prediction, this would yield the best `theoretical' prediction
for the normalization in the lower $M_{2l}$ mass region of interest.
But, bringing the systematic error in the `theoretically'
computed background normalization
below the critical 10\% level seems quite problematical.

Further, there would remain the question of detector efficiency and such
as a function of lepton momentum (which feeds into detector efficiency
for a given $M_{2l}$). This distorts the theoretical expectations
so as to reduce the accuracy for the above procedure.
Aside from detection efficiencies,
isolation cuts could be $p_T$-dependent and perhaps hard to understand.
These would be critical questions for the experimental groups.  Clearly,
they would have to understand their detector(s) very well. Reducing
uncertainties from this source to something like
the 5\% level is possibly achievable in a mature well-studied detector.
\Ref\einsweiler{K. Einsweiler, private communication.}

If data were available, one would take
the signal shape, estimate some uncertainties coming from production model
variations, then do the same for the background, and try fitting the data
with these variations, and assess the significance of an ``excess''.
This would not be an easy job, and it would be difficult
to claim discovery of the source of EWSB as a 10\% excess
in a $\sim 50$ GeV wide region sitting on top of a complex background.
In addition, it is entirely possible that the necessity of employing
a low threshold in $p_T^l$ would allow other detector dependent backgrounds
to creep into the `0-jets' $M_{2l}$ distributions.
Thus, the observability of the $\hsm$ in the $2l$ mode seems quite
questionable. At best it can be noted
that hints of an $\hsm$ event excess begin to emerge
for $\mhsm \gsim 140\gev$, where $S/B$ exceeds 10\%. For the best case
of $\mhsm\sim 170\gev$, the signal could possibly be detected given that
$S/B$ has reached a level of order 20(30)\% for $p\anti p$($pp$) collisions.
However, by $\mhsm=200\gev$ the anticipated systematic uncertainties will
probably prohibit seeing even a hint of the $\sim 8\%$
signal event excess, given even an optimistic
assumption as to the accuracy with which
the normalization of the $WW$ continuum background will be determined.
Due to larger $S/B$ values (not to mention
higher instantaneous luminosity), a $pp$ machine
would have a clear advantage in searching for an excess of $2l2\nu$ events.

\FIG\llnunu{
The $gg\rta \hsm\rta ZZ\rta ll \nu\nu$ ($l=e,\mu$) signal and
$q\anti q \rta ZZ \rta ll \nu\nu$ background event rates as a function of the
transverse mass, $M_T$, in 5 GeV bins.  Signals for
$\mhsm=170$, $200$, $230$, $270$, $300$ and $400\gev$ are shown.
QCD `K' factors of 1.1 and 1.5 have been included in the $ZZ$
continuum background and $gg$ fusion signal, respectively.
Optimistic lepton resolution/acceptance choices, case ii)-b), are employed.
}
\midinsert
\vbox{\phantom{0}\vskip 5.0in
\phantom{0}
\vskip .5in
\hskip -0pt
\special{ insert scr:llnunu_4tev.ps}
\vskip -1.45in }
\centerline{\vbox{\hsize=12.4cm
\Tenpoint
\baselineskip=12pt
\noindent
Figure~\llnunu:
The $gg\rta \hsm\rta ZZ\rta ll \nu\nu$ ($l=e,\mu$) signal and
$q\anti q \rta ZZ \rta ll \nu\nu$ background event rates as a function of the
transverse mass, $M_T$, in 5 GeV bins.  Signals for
$\mhsm=170$, $200$, $230$, $270$, $300$ and $400\gev$ are shown.
QCD `K' factors of 1.1 and 1.5 have been included in the $ZZ$
continuum background and $gg$ fusion signal, respectively.
Optimistic lepton resolution/acceptance choices, case ii)-b), are employed.
}}
\endinsert

 \TABLE\llnunutable{}
 \topinsert
 \titlestyle{\twelvepoint
 Table \llnunutable: For $ZZ\rta ll \nu\nu$,
we tabulate as a function of Higgs boson mass: the signal and background
rates $S$ and $B$, for $p\anti p$ ($pp$),
for $L=30\fbi$ summed over a 10 GeV interval; the associated
$S/\sqrt B$ values, as the
absolute upper bound on the observability of the signal; and
the $L$ (in $\fbi$) required for a $S/\sqrt B=5$ signal level.
Optimistic lepton resolution/acceptance choices, case ii)-b), are employed.
}
 \bigskip
\def\tstrut{\vrule height 14pt depth 4pt width 0pt}
 \thicksize=0pt
 \hrule \vskip .04in \hrule
 \begintable
 $\mhsm$ | $S$ | $B$ | $S/\sqrt B$ | $L(5\sigma)$ \cr
 200 | 144 | 390(330)   | 7.3(8.0) | 14(12)       \cr
 230 | 66  | 132(78)    | 5.6(7.3) | 25(14)      \cr
 270 | 29  | 54(30)     | 3.8(5.2)  | 52(28)     \cr
 300 | 21  | 21(13)     | 4.2(5.7)   | 40(22)
\endtable
 \hrule \vskip .04in \hrule
 \endinsert

\vskip .05in
\noindent\undertext{$gg\rta\hsm\rta ZZ\rta ll \nu\nu$}
\vskip .05in

For $\mhsm\gsim 2\mz$, it is also worth examining the $\hsm\rta ZZ\rta
ll \nu\nu$ mode, in which one of the $Z$'s decays to
neutrinos.\refmark{\zzllll}
The main irreducible background arises, of course, from $ZZ$ continuum
production.  Several variables can be used to reveal the Higgs boson mass
peak.  Here we have chosen to employ the transverse mass defined by
$M_{T}\equiv 2\sqrt{\mz^2+p_{T\,Z}^2}$.
Although rather nice mass peaks are revealed in Fig.~\llnunu\ and event rates
are reasonable, there will certainly be additional
backgrounds, as discussed below.  As in previous channels, we first
tabulate nominal $S$ and $B$ values and associated
$L=30\fbi$ $S/\sqrt B$ statistical significance, ignoring the
additional backgrounds. Focusing on the two largest 5 GeV bins (which
yields the best results) we obtain the results in Table~\llnunutable.
At $\mhsm=200$ and $230\gev$ acceptable values for $S/\sqrt B$
appear.  However, this ignores the possibly large reducible backgrounds
from processes such as $Zg$ production in which the $g$
occasionally produces very little visible energy in the detector.
The $Zg$ process has a very high event rate and could lead to
a background for this signal if the detector does not have
large rapidity coverage and few cracks, \etc\
At the SSC, the SDC detector studies\refmark\sdctdr\ found that
these backgrounds were so large at low transverse hadronic energies
that a Higgs boson with mass below about $500\gev$ could not
be detected in this way.  Presumably the backgrounds are somewhat
less severe at the lower $\sqrt s=4\tev$ energy of interest here.
However, our expectation is that they will probably render
this mode useless.  Nonetheless, given the promising level of
the nominal $S/\sqrt B$ values obtained without including these
backgrounds, it would clearly be worthwhile to pursue this issue.

\vskip .05in
\noindent\undertext{$gg\rta\hsm\rta ZZ\rta 2ljj$}
\vskip .05in

The $\hsm\rta ZZ\rta 2ljj$ channel yields signal
and event rate distributions  in the $M_{2ljj}$ mass that are
somewhat narrower than those obtained using $M_{ljj}^{cluster}$
in the $WW$ channel.  However, the event rates in the $2ljj$ channel
are about a factor of $6$ lower. Meanwhile, the $Zjj$ and $ZZ$
continuum backgrounds have about the same relative size as in the
$WW$ case.  The largest statistical significance is $\nsd\sim 4.8$
for $L=30\fbi$ at $\mhsm=170\gev$ , for the optimistic detector/cuts case.
Keeping in mind the additional systematic errors not included in this $\nsd$
estimate, this channel does not appear to be useful.

\vskip .05in
\noindent\undertext{$gg\rta\hsm\rta \gam \gam$}
\vskip .05in

For the inclusive $\gam\gam$ channel,
we find that even if the resolution is such that
the entire Higgs boson signal is contained within a 1 GeV bin, which give
a signal rate of about 100 events for $100 \lsim \mhsm \lsim 140\gev$,
the statistical significance $\nsd$, computed as $\nsd=S/\sqrt B$,
is never much above 1 due to the overwhelming background from the
$q \anti q \rta \gam\gam$ continuum. This channel does not appear to be useful
at a 4 TeV Tevatron.

\vskip .05in
\noindent\undertext{$gg\rta \hsm\rta \tau^+\tau^- \rta 2l 4\nu$}
\vskip .05in

Finally, one may consider the possibility of using the
$\hsm\rta \tau^+\tau^- \rta 2l 4\nu$ channel, since
the branching fraction for $\hsm \rta \tauptaum$ can be as high as
4\% for $\mhsm \lsim 140\gev$, and the Higgs mass
peak could be reconstructed if we require some finite transverse
momentum for the $\tau$ pair.\Ref\tautau{R.K. Ellis, I. Hinchliffe, M. Soldate,
and J. van der Bij, \npbj{297} (1988) 221.}
Unfortunately, the Drell-Yan backgrounds
$\gam^*, Z \rta \tau^+\tau^-$ and $l \anti l$, as well as
$\wp\wm\rta 2l 2\nu$ are so overwhelming that there is little hope
to extract the signal.

\vskip .15in
\noindent{\bf 3. Conclusion}
\vskip .075in

We have studied the ability of an upgraded Tevatron with  $\sqrt s=4\tev$
to search for a Standard Model Higgs boson with mass beyond
the reach of LEP-200. Since such an upgrade would be most useful if
detection were possible prior to the full luminosity operation
of the LHC, we have employed an integrated luminosity of $L=30\fbi$
in our evaluations of discovery potential.

At low $\mhsm$, the $Wb\anti b$ mode provides clear signals,
\Ref\willenetal{Similar conclusions have been reached in a recent work
by A. Stange, W. Marciano, and S. Willenbrock (to appear as a BNL preprint).}\
especially if double $b$-tagging is employed.
However, this mode cannot be pushed beyond $\mhsm\sim 120\gev$,
and is most likely restricted to $\mhsm\lsim 110\gev$. Thus, other
modes must be considered for $\mhsm\gsim 110\gev$.

In the very clean $ZZ\rta 4l$ mode the feasibility of detecting the $\hsm$
is clearly limited by the small event rate. However,
at $\mhsm=150\gev$ one expects about 13 events (with negligible
background) and at $\mhsm=200\gev$ the signal and background rates
yield $S/\sqrt B=28/\sqrt{28}\sim 5$ for $L=30\fbi$, before
including track-finding efficiency. An $\hsm$ with mass as
large as $230-250\gev$ yields a $3\sigma$ level signal
which might be adequate for discovery given the cleanliness of this mode.
However, to achieve the above rates requires that leptons with transverse
momentum down to $10\gev$ be retained.

For $\mhsm$ between about $150\gev$ and $200\gev$
$\hsm$ detection in the $WW\rta l\nu jj$ channel would also be difficult.
For instance, the $\mhsm=170\gev$ signal reaches a nominal
$S/\sqrt B$ of order 10, but might not be easy to
detect given the systematic uncertainties associated
with a signal to background ratio of $S/B\sim 0.01$ and
a background that peaks in this same mass region.

In the $WW\rta 2l 2\nu$ mode,
event rates and/or systematics will certainly prevent
detection of the $\hsm$ in the $110-140\gev$ mass range.
For $140\lsim \mhsm\lsim 180\gev$, where $S/B$ exceeds 10\%,
there is a remote chance that systematics problems could be overcome
and a broad event excess due to the $\hsm$ distinguished.
However, as discussed, this should be regarded as very borderline.
The need for employing low thresholds for accepting leptons makes this $2l$
mode especially detector-dependent.

In the $\hsm\rta ZZ\rta ll \nu\nu$ channel,
the Higgs boson signals exhibit a decent ($\gsim 5\sigma$) nominal statistical
significance with respect to the $ZZ$ continuum background
for $2\mz \lsim\mhsm\lsim 250\gev$.  However, we are very concerned
that the signal will be swamped by large rate $ll+jets$
backgrounds that have a small tail where the jets end up depositing
only a small amount of transverse hadronic energy in the detector.
A particularly problematical example is $gZ$ production (where the $g$ leaves
little trace in the detector, \eg\ goes down the beam line).

Finally, for $110-120\lsim \mhsm\lsim 140-150\gev$ and $\mhsm\gsim 250\gev$
there is little hope of detecting the $\hsm$ at a $\sqrt s=4\tev$
Tevatron upgrade.

We emphasize that to obtain potential signals for Higgs boson masses
beyond the reach of LEP-200 will require multiple years of running at
$L=10\fbi$ per year, even in the more favored
$\lsim 110-120\gev$ and $150-230\gev$ mass ranges.
Overall, we do not think that a 4 TeV upgrade of the Tevatron can
be justified on the basis of its potential for Standard Model Higgs
boson discovery.

Believers in the minimal supersymmetric model (MSSM) will note that if the
CP-odd scalar has mass $\mhsm\gsim 2\mz$, then the light CP-even scalar, the
$\hl$, will have relatively SM-like couplings.  Meanwhile, its mass, even
after radiative corrections would certainly be below about
$160\gev$,
\Ref\hmass{For a review, see H.E. Haber,
{\it Perspectives in Higgs Physics}, ed. G. Kane,
World Scientific Publishing (1992).}\
and most probably (\ie\ for stop squark mass below about 500 GeV)
would lie in the $\mhl\lsim 140\gev$ region.  That is the $\hl$
mass may well reside in exactly the region of greatest weakness for
a 4 TeV Tevatron. Even for $\mhl<100-110\gev$ there could be a
problem due to the possibly present
$\hl\rta I$ invisible decay modes, such as $I=\cnone\cnone$ ---
where $\cnone$ is the lightest supersymmetric particle.
Further investigation is needed to determine if there is a detectable
signal in the associated $W\hl\rta l \ptmiss$ production/decay mode.
Meanwhile, the other MSSM Higgs bosons are almost certainly undetectable
at this minimally upgraded machine.

\smallskip\centerline{\bf Acknowledgements}
\smallskip
We would like to thank W. Marciano, A. Stange and S. Willenbrock
for discussions regarding several of the modes considered here.
We are also indebted to K. Einsweiler for numerous consultations
on experimental and analysis issues.
This work has been supported in part by Department of Energy
grant \#DE-FG03-91ER40674,
and by Texas National Research Laboratory grant \#RGFY93-330.
T.H. is supported in part by a UC-Davis Faculty Research Grant.

\smallskip
\refout
\end
\bye